\documentclass[lettersize,journal, table, xcdraw]{IEEEtran}
\usepackage{amsmath,amsfonts}
\usepackage{algorithmic}
\usepackage{algorithm}
\usepackage{array}
\usepackage{caption, subcaption, graphicx, svg}
\usepackage{textcomp}
\usepackage{stfloats}
\usepackage{url}
\usepackage{verbatim}
\usepackage{graphicx}
\usepackage{cite}
\usepackage{xspace}
\usepackage[colorlinks]{hyperref}
\hypersetup{
    citecolor=red,
    linkcolor=blue
}
\usepackage{multirow}

\begin{document}

\title{Vectorized Sequence-Based Chunking for Data Deduplication}

\author{
\IEEEauthorblockN{Sreeharsha Udayashankar}
\IEEEauthorblockA{\textit{University of Waterloo} \\
}
\and
\IEEEauthorblockN{Samer Al-Kiswany}
\IEEEauthorblockA{\textit{University of Waterloo} \\
\{s2udayas, alkiswany\}@uwaterloo.ca}
}

\definecolor{BrickRed}{HTML}{B6321C}
\definecolor{RoyalBlue}{HTML}{4169E1}

\newcommand{\sysname}{SeqCDC\xspace}
\newcommand{\samer}[1]{\textcolor{green}{\textbf{[Samer: #1]}}}
\newcommand{\harsha}[1]{\textcolor{BrickRed}{\textbf{Harsha: #1}}}


\maketitle

\begin{abstract}
Data deduplication has gained wide acclaim as a mechanism to improve storage efficiency and conserve network bandwidth. Its most critical phase, data chunking, is responsible for the overall space savings achieved via the deduplication process. However, modern data chunking algorithms are slow and compute-intensive because they scan large amounts of data while simultaneously making data-driven boundary decisions.

We present \sysname, a novel chunking algorithm that leverages lightweight boundary detection, content-defined skipping, and SSE/AVX acceleration to improve chunking throughput for large chunk sizes. Our evaluation shows that \sysname achieves $15\times$ higher throughput than unaccelerated and $1.2\times$--$1.35\times$ higher throughput than vector-accelerated data chunking algorithms while minimally affecting deduplication space savings.

\end{abstract}

\begin{IEEEkeywords}
    Data storage, Data deduplication, SIMD, Cloud computing 
\end{IEEEkeywords}

\section{Introduction}

Data generation rates have skyrocketed in recent years, leading to the explosion of the amount of data stored on the cloud~\cite{data_growth}. Cloud storage providers employ numerous mechanisms to deal with this data influx, such as distributed file systems~\cite{google_file_system, hdfs}, novel storage architectures~\cite{raid_1994, NAS_2000}, data compression~\cite{data_compression_survey, data_compression_book} and data deduplication~\cite{past_future_dedup, demystifying_dedup}.

Data deduplication has been widely employed in production by cloud storage providers such as Microsoft~\cite{dedup_intro}, EMC~\cite{emc_backup} and IBM~\cite{ibm_dedup}. A large percentage of the data stored by these providers is redundant~\cite{dedup_intro}. Data deduplication helps identify and eliminate these redundant portions, reducing storage costs by up to 80\% ~\cite{emc_backup, venti}. Deduplication is performed at the \textit{chunk-level}, after dividing files into \textit{chunks}~\cite{dedup_techniques}. 

The division of files into chunks is achieved using data chunking algorithms~\cite{lbfs}, which dictate the space savings achieved by the deduplication system as a whole. Data chunking algorithms fall into two categories: fixed-size and content-defined chunking (CDC). Fixed-size chunking divides files into chunks of an equal pre-specified size. While this approach has been used by traditional backup systems such as Venti \cite{venti} and OceanStore \cite{oceanstore}, it achieves poor space savings due to its vulnerability to insertions and deletions that cause bytes to shift, i.e., \textit{byte-shifting}~\cite{lbfs}. 

To mitigate byte-shifting, modern deduplication systems instead resort to Content-Defined Chunking (CDC) algorithms~\cite{ae, sscdc, fastcdc, gear_hash, lbfs, ram, tttd_hp}. These algorithms make data-driven boundary decisions using the file's contents, effectively handling the byte-shifting problem. 
Numerous data chunking algorithms are in use today and can be broadly divided into hash-based~\cite{fastcdc, lbfs, tttd_hp, gear_hash} and hashless algorithms~\cite{ae, ram, maxp}. Hash-based algorithms use rolling hash functions to derive chunk boundaries, while hashless algorithms treat each byte as a value and derive chunk boundaries using conditions based on local minima or maxima. Note that in either case, a \textit{fingerprint} is generated using collision-resistant hash algorithms~\cite{sha256} after a chunk boundary is identified.

CDC algorithms suffer from four limitations that impact their throughput. First, they rely on expensive rolling hash functions \cite{lbfs, gear_hash} and minima-maxima searches \cite{ae, ram} to determine chunk boundaries. Second, they sequentially scan the ingested data stream. This reduces throughput and increases end-to-end processing time, as real systems store terabytes of data. Third, they fail to utilize the SIMD capabilities of modern CPUs to accelerate data processing. Finally, they are designed to target datasets that benefit from smaller chunks of size $512$\,B -- $4$\,KB. Such small chunks increase metadata overhead~\cite{past_future_dedup} and impact system throughput due to the random access and frequent transfer of small chunks. 

State-of-the-art CDC techniques aimed to solve some of these shortcomings. FastCDC \cite{fastcdc} replaced the expensive Rabin's hash algorithm \cite{lbfs} with Gear hashing \cite{gear_hash} to reduce boundary-detection overhead. AE \cite{ae} and RAM \cite{ram} reduce the overhead by avoiding hashing entirely, instead relying on minima/maxima to determine chunk boundaries. SS-CDC \cite{sscdc} and our previous work, VectorCDC \cite{vectorcdc}, accelerate CDC algorithms with SSE/AVX instructions \cite{vector_inst_support} offered by modern CPUs. However, these studies only address a few specific limitations and fail to do so effectively.

We present \sysname, a novel CDC algorithm that comprehensively addresses the limitations of modern CDC algorithms. \sysname uses three optimizations to improve chunking throughput: \textit{lightweight boundary judgment}, \textit{content-based data skipping}, and \textit{vector acceleration}. Lightweight boundary judgment reduces boundary detection overhead by using monotonically increasing/decreasing sequences, avoiding complex hashing and minima-maxima searches (\S\ref{sec:design-boundary-judgement}). To avoid scanning the entire source data, \sysname skips scanning selective data regions. However, to minimize the impact on deduplication efficiency, data skipping is regulated using content-based heuristics i.e. content-based skipping (\S\ref{sec:design-skip}). \sysname has been designed with vector acceleration in focus, and uses SSE/AVX instructions \cite{vector_inst_support} to improve chunking throughput (\S\ref{sec:design-vector}). Finally, \sysname scales its throughput with chunk size, i.e., it offers higher throughput at the larger chunk sizes favored by deduplication systems (\S\ref{sec:eval_chunking_speed}). 

Our evaluation compares \sysname to seven unaccelerated and three vector-accelerated chunking algorithms using a variety of real world datasets (\S\ref{sec:evaluation}). We show that \sysname improves chunking throughput by $10\times$ over unaccelerated chunking algorithms and $1.25\times$--$1.35\times$ over vector-accelerated CDC algorithms, while achieving comparable deduplication space savings. Our code is publicly available with DedupBench\footnote{\url{https://github.com/UWASL/dedup-bench}}~\cite{dedupbench}. 



\section{Background and Motivation}

\label{sec:bg}

Data deduplication~\cite{dedup_intro, lbfs} is used by cloud storage providers to detect duplicate data, allowing them to eliminate the costs associated with its storage and transmission. Data deduplication consists of the following steps~\cite{past_future_dedup}:

\begin{itemize}
    \setlength\itemsep{0.5 em}
    \item \textit{File Chunking:} Splitting a file into chunks using a data chunking algorithm is one of the primary steps in data deduplication. Deduplicating these chunks provides more space savings than file-level deduplication~\cite{lbfs}.
    \item \textit{Chunk Hashing:} Each data chunk is hashed using a collision-resistant hashing algorithm, such as SHA-256~\cite{sha256} or MD5~\cite{md5}, to obtain a fingerprint.
    \item \textit{Fingerprint Comparison:} The fingerprint is compared against a database of previously observed fingerprints. A duplicate fingerprint, i.e., one observed before, indicates an underlying duplicate chunk, which can be eliminated.
    \item \textit{Data Storage:} Non-duplicate data chunks are saved on the storage medium, and their fingerprints are added to the fingerprint database.
\end{itemize}

Chunking is a critical part of this pipeline; it occurs on the critical path during data uploads and directly impacts the overall space savings and throughput associated with deduplication systems. Space savings represent the total space conserved by using deduplication. 

\begin{equation}
    Space~savings = \frac{Original~Size - Deduplicated~Size}{Original~Size} 
\end{equation}

The size of the fingerprint database is tied to the average chunk size. Smaller average sizes lead to more chunks and more fingerprints, thus increasing the size of the database and associated fingerprint comparison overheads. To minimize this overhead, deduplication systems in production tend to favor larger chunk sizes.

\subsection{Content-Defined Chunking (CDC) Algorithms}
\label{sec:bg_chunking}

While fixed-size chunking has been used in the past \cite{venti}, it remains vulnerable to byte shifting and achieves poor space savings \cite{lbfs}. Content-Defined Chunking (CDC) algorithms instead make chunk boundary decisions based on the data contents. Numerous CDC algorithms~\cite{ae, fastcdc, sscdc, gear_hash, lbfs, ram, maxp, tttd_hp} have been proposed for data deduplication. 

These algorithms slide a fixed-size window over the data within the source file. When the window's data meets pre-specified conditions, they insert a \textit{chunk boundary} at the end of the window. By repeating this across the entire file, they divide it into data chunks. Each CDC algorithm has parameters that can be tuned to change the average size of generated chunks. CDC algorithms can be classified into hash-based and hashless algorithms \cite{dedupbench}. 

Hash-based chunking algorithms, such as Rabin's chunking~\cite{lbfs}, insert chunk boundaries only when the hash value of the window's data matches a pre-specified mask. The hashing algorithms used here are typically not collision-resistant.
For instance, with Rabin's Chunking (RC)~\cite{lbfs}, chunk boundaries are inserted when the lower order 13 bits of the hash value are zero. While Rabin's chunking achieves high deduplication ratios, it is very slow. TTTD~\cite{tttd_hp} uses Rabin's hashing but improves space savings by simultaneously checking for two hash value conditions. The secondary condition is always computed but only used if a boundary is not found beyond a pre-specified size with the primary condition.  

The CRC~\cite{sscdc} and Gear ~\cite{gear_hash} hashing functions have lower overheads than Rabin's hashing. FastCDC (FCDC)~\cite{fastcdc} uses Gear hashing and implements two optimizations to improve chunking throughput: sub-minimum skipping and chunk size normalization. Sub-minimum skipping skips scanning data up to the minimum chunk size at the beginning of each chunk. Chunk size normalization dynamically relaxes the boundary condition to ensure that generated chunks are close to the expected average chunk size. 

Hashless algorithms such as Asymmetric Extremum (AE)~\cite{ae} and Rapid Asymmetric Maximum (RAM)~\cite{ram} also slide fixed-size windows over the source data. AE attempts to identify a window such that the starting byte's value is greater than all the bytes before it and not less than the other bytes within the window. When such a window is found, AE inserts a chunk boundary at the end of the window. AE is $4-5\times$ faster than Rabin's chunking. Rapid Asymmetric Maximum~\cite{ram} inserts chunk boundaries when the byte immediately outside the window has a value greater than the maximum-valued byte within the window. These algorithms avoid hashing when determining chunk boundaries, reducing boundary-detection overhead and achieving high chunking throughput \cite{lowentropy, vectorcdc}.



\begin{figure}[t]
\centering
    \begin{subfigure}{0.45\linewidth}
        \centering
        \includegraphics[width=\linewidth]{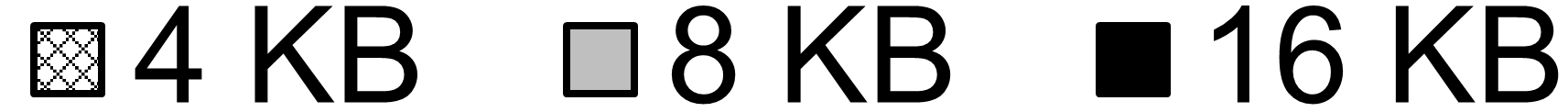}
    \end{subfigure}
    \hspace*{\fill}\\
    \begin{subfigure}{0.8\linewidth}
    \centering
        \includesvg[inkscapelatex=false, width=\linewidth]{figures/chunking_speed_motivation.svg}
    \end{subfigure}
    \caption{Chunking throughput on randomized data}
    \label{fig:chunking_speed_motivation_cdc}
\end{figure}

\subsection{Chunking Throughput Analysis}
\label{sec:motivation_chunking_speed}

As noted above, deduplication systems in production prefer a smaller number of larger size chunks, to minimize fingerprinting overheads. However, state-of-the-art CDC algorithms are designed with smaller chunk sizes in mind i.e., their throughput remains constant across chunk sizes.

Figure \ref{fig:chunking_speed_motivation_cdc} shows the throughput achieved by AE \cite{ae}, CRC \cite{sscdc}, FastCDC (FCDC) \cite{fastcdc}, Gear-based chunking \cite{gear_hash}, RAM \cite{ram}, Rabin's Chunking (RC) \cite{lbfs} and TTTD \cite{tttd_hp} when chunking a 1GB file containing random data.We compare the chunking throughput of these algorithms across three average chunk sizes: 4\,KB, 8\,KB, and 16\,KB. This experiment was run on a machine with an Intel Icelake CPU, the details of which are in \S\ref{sec:evaluation}.

We note that the throughputs of all these algorithms do not scale with chunk size, as they process the entire data stream regardless of target size. One of the motivating factors behind \sysname was to achieve higher throughput at larger chunk sizes used in production. 

When targeting larger chunk sizes, most of the scanned data does not qualify as chunk boundaries. To minimize wasteful computation at larger chunk sizes, we propose skipping data regions during scanning. As random skipping can severely degrade the space savings achieved by deduplication, we use \textit{content-defined data skipping} to skip scanning only certain regions (\S\ref{sec:design-skip}) using data-based heuristics. At larger chunk sizes, \sysname can afford to skip larger amounts of data without impacting space savings, leading to higher throughput.


\subsection{Accelerating CDC algorithms with vector instructions}
\label{sec:bg-vector}

Vector instruction sets \cite{vector_inst_support} are supported by most modern Intel and AMD CPUs. These instructions allow for the execution of arithmetic/logical operations simultaneously on multiple pieces of data, i.e., the \textit{Single-Instruction Multiple-Data (SIMD)} paradigm. To do this, they rely on special vector registers provided within the CPU, packing multiple values into them and operating on all the values with a single operation such as an addition or subtraction. Depending on the amount of data they process at once, these instructions can be classified as SSE-128, AVX-256, and AVX-512, i.e., 128-bit, 256-bit, or 512-bit. While AVX-512 instructions are only supported by the newest Intel and AMD CPUs, SSE-128 and AVX-256 support has been available since 2003 and 2011, respectively.

Vector instructions have previously been used to accelerate mathematical operations \cite{vector_matmul, vector_sort} and multimedia applications \cite{vector_mult}. SS-CDC \cite{sscdc} previously attempted to accelerate hash-based CDC algorithms, such as CRC-32 and Gear \cite{gear_hash}, using AVX-512 instructions. They decouple the rolling hash and boundary detection phases to accelerate them separately. However, many hash-based algorithms such as FastCDC \cite{fastcdc} and TTTD \cite{tttd} use minimum chunk size skipping to improve throughput. Running the rolling hash phase on the entire source data and identifying boundaries later in a separate phase eliminates this throughput benefit. Additionally, due to the dependency between adjacent bytes when calculating hash values, accelerating rolling hash algorithms is complicated. To solve this, SS-CDC resorted to processing different regions of the data simultaneously with AVX-512 instructions, i.e, rolling with multiple heads. As this requires expensive vector \texttt{scatter/gather} instructions, the speedups achieved are limited, as shown by our previous work VectorCDC \cite{vectorcdc}.

VectorCDC \cite{vectorcdc} instead accelerates hashless algorithms such as AE \cite{ae} and RAM \cite{ram} with vector instructions. It identifies two phases common to these algorithms, the \textit{Extreme Byte Search} and \textit{Range Scan} phases, accelerating them using different vector-based techniques. Using these techniques, VectorCDC \cite{vectorcdc} achieved orders of magnitude higher speedup for hashless algorithms than what SS-CDC \cite{sscdc} achieved for hash-based ones. While beneficial, vector-acceleration alone cannot improve chunking throughput at larger sizes (\S\ref{sec:eval_chunking_speed}).

Despite being hashless, \sysname does not follow the same paradigm as AE and RAM. \sysname uses content-defined skipping (\S\ref{sec:design-skip}). Additionally, instead of relying on minimum/maximum byte values to detect chunk boundaries, \sysname uses monotonically increasing/decreasing sequences (\S\ref{sec:design}). Thus, VectorCDC's approach is incompatible with \sysname. 


\section{\sysname's Design}
\label{sec:design}

\sysname is designed to insert chunk boundaries when fixed-length sequences of monotonically increasing/decreasing bytes are detected. \sysname can operate in either \texttt{Increasing} mode i.e. targeting increasing order sequences or \texttt{Decreasing} mode. Note that these two modes are exclusive of each other. Figure \ref{fig:seqcdc_example} shows an example of \sysname's operation. \S\ref{sec:byte shifting} discusses how \sysname is resistant to byte shifting.


\sysname utilizes three parameters: \textit{SeqLength}, \textit{SkipTrigger} and \textit{SkipSize}, each described in detail in the following subsections. \sysname includes four optimizations we discuss in detail: lightweight boundary detection, ignoring data at the beginning of a chunk, content-based data skipping, and acceleration with vector instructions.

\subsection{Lightweight Boundary Detection}
\label{sec:design-boundary-judgement}


To avoid complex hashing operations, \sysname treats each byte within the data stream as an independent value similar to existing hashless CDC algorithms \cite{ae, ram}. However, to improve chunking throughput even further, \sysname reduces the overheads associated with boundary detection by avoiding minima/maxima searches. 

Instead, \sysname looks for \textit{fixed-length monotonically increasing/decreasing} sequences of bytes
and inserts chunk boundaries whenever such a sequence is detected. The sequence must have a length of \textit{SeqLength} to be considered a boundary candidate. Once such a sequence is found, a boundary is inserted at its end.

\textbf{Modes of operation.} \sysname can be used in \texttt{Increasing} or \texttt{Decreasing} mode. Both these modes are exclusive of each other. While in  \texttt{Increasing} mode, \sysname targets monotonically increasing sequences. On the other hand, it targets monotonically decreasing sequences in \texttt{Decreasing} mode. Depending upon dataset characteristics, one mode may be more effective than the other. 

\begin{figure}[t]
    \centering
    \includegraphics[scale=0.43]{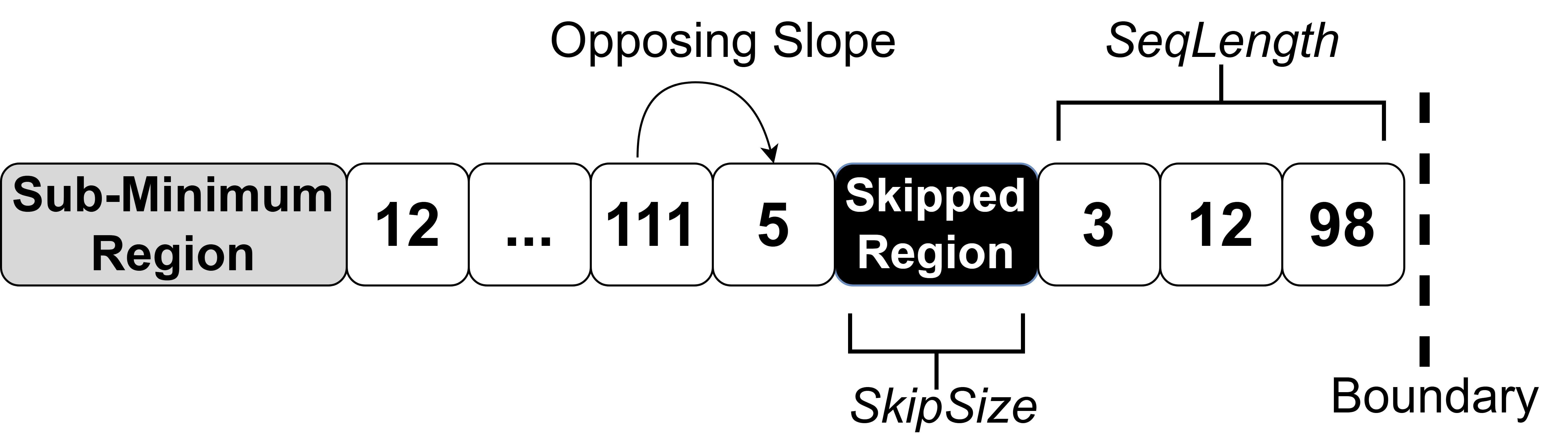}
    \caption{An example of a chunk generated by \sysname}
    \label{fig:seqcdc_example}
\end{figure}

Figure \ref{fig:seqcdc_example} shows an example of \sysname operating in \texttt{Increasing} mode with a \textit{SeqLength} of 3. A chunk boundary is inserted after the byte with value 98, as it forms an increasing sequence with the bytes preceding it.

\subsection{Ignoring Sub-minimum Regions}

\sysname utilizes the concept of ignoring data at the beginning of each chunk introduced within previous literature~\cite{tttd, fastcdc} to increase chunking throughput. \sysname skips scanning data of size (\textit{minimum\_chunk\_size} - \textit{SeqLength}) at the beginning of each chunk (\texttt{"Sub Minimum Region"} in Figure \ref{fig:seqcdc_example}). 
 
Increasing the minimum chunk size allows \sysname to skip over larger amounts of data at the beginning of each chunk, increasing chunking throughput. However, when performed excessively, this may negatively impact space savings on some datasets. The minimum chunk size for \sysname is 25-50\% the average chunk size, similar to existing algorithms~\cite{tttd, fastcdc}.  

\subsection{Content-based Data Skipping}
\label{sec:design-skip}

\sysname additionally improves chunking throughput by skipping scanning certain data regions when looking for chunk boundaries (\texttt{"Skipped Region"} in Figure \ref{fig:seqcdc_example}). Randomly skipping data regions can lead to missed boundaries, lowering byte-shifting resistance and negatively affecting space savings. To avoid this, \sysname adopts a novel content-based data skipping mechanism, i.e., data regions are skipped over only when skip conditions are met.


\sysname skips scanning data within \textit{unfavorable regions}, i.e., data regions with byte sequences in an order opposing the target sequence. For instance, when in \texttt{Increasing} mode, regions with decreasing order sequences are considered unfavorable. When \textit{SkipTrigger} pairs of bytes in opposing order are detected, \sysname decides that the current region is unfavorable and skips scanning the next few bytes in the hope of landing in a favorable region, with bytes in the chosen order. For instance, in Figure ~\ref{fig:seqcdc_example}, the skip condition is triggered after the byte with a value of 5, causing the next \textit{SkipSize} bytes to be ignored. The \textit{SkipSize} is kept small at 256-512 bytes, to avoid skipping over large sections of data. After a skip is triggered, \sysname resets its counters and resumes scanning for boundaries. 

Larger \textit{SkipSizes} improve chunking throughput. While larger \textit{SkipSizes} are feasible for larger chunks, they may result in a disproportionately high amount of data skipped within smaller chunks, negatively affecting space savings. \sysname overcomes this by adjusting the \textit{SkipSize} from $256-512$ bytes, depending on the expected average chunk size.

Data skipping can potentially impact byte-shifting resistance. \sysname therefore trades off a small reduction in space savings for higher chunking throughput. Section \ref{sec:byte shifting} discusses this trade-off in greater detail. Additionally, in our evaluation  (\S\ref{sec:eval_space_savings}), we show that this design decision minimally impacts space savings in real datasets. 




\subsection {Accelerating \sysname with vector instructions}
\label{sec:design-vector}

\begin{figure}[t]
    \centering
    \includegraphics[width=\linewidth]{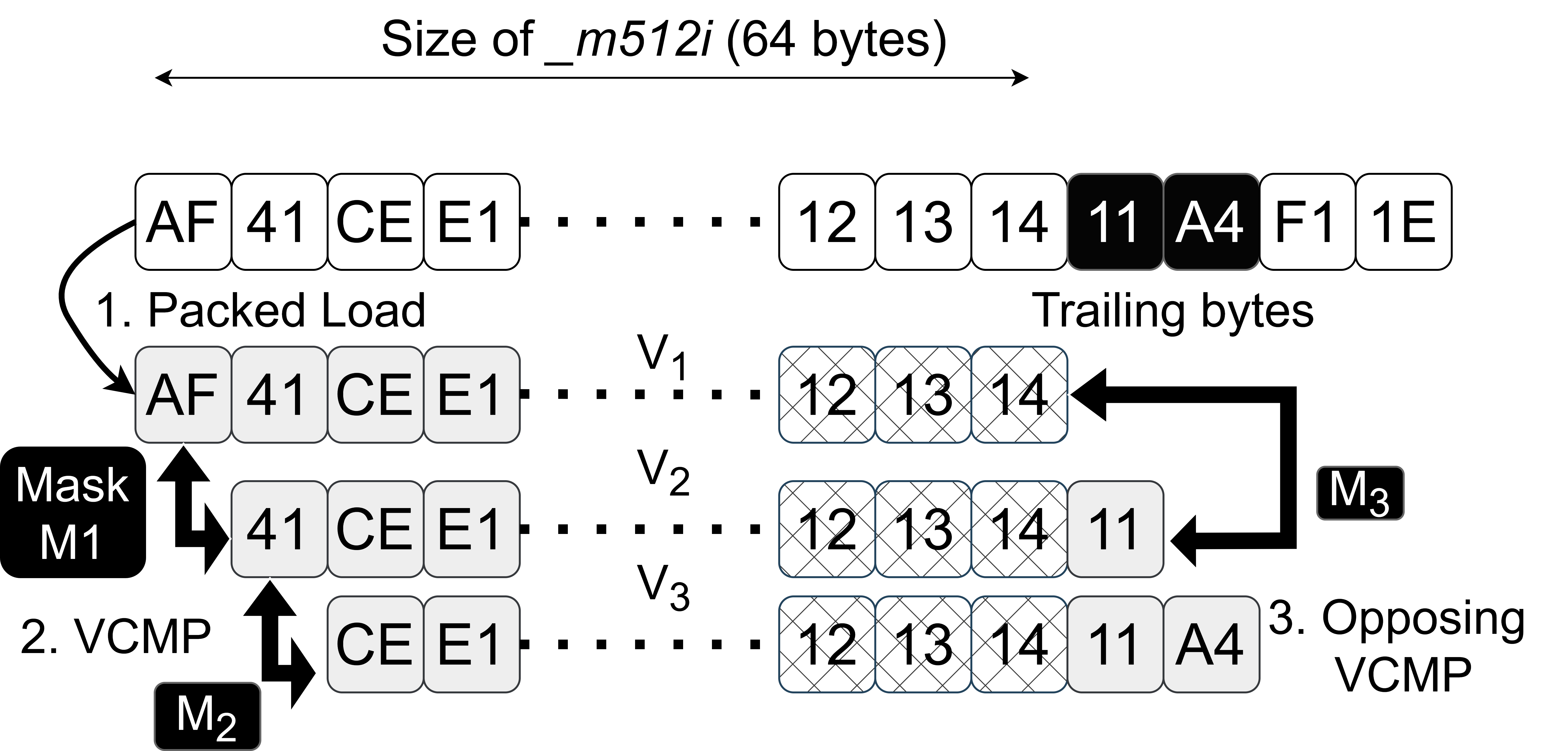}
    \caption{Accelerating \sysname with AVX-512 instructions}
    \label{fig:seqcdc_vector}
\end{figure}

VectorCDC \cite{vectorcdc} demonstrated the throughput benefits that can be obtained by using vector instructions to accelerate data chunking algorithms. However, VectorCDC's approach cannot be directly applied to accelerate \sysname (\S\ref{sec:bg-vector}), as \sysname relies on detecting monotonically increasing/decreasing sequences for chunk boundaries. We propose an alternate vector-based method to accelerate \sysname in this section.

Figure \ref{fig:seqcdc_vector} shows an example of accelerating \sysname operating with a \textit{SeqLength} of 3 and \texttt{Increasing} mode using AVX-512 instructions \cite{vector_inst_support}. The figure shows a byte range with byte values \texttt{AF}--\texttt{1E} that need to be scanned for boundaries. A chunk boundary sequence starting at byte \texttt{12} exists in this region and is shown with a cross-stitched pattern. Let us assume scanning begins at byte \texttt{AF}.

\textbf{Scan Procedure.} We start by loading the 64 bytes \texttt{AF}--\texttt{14} into a vector register $V_1$ in packed fashion (Step 1). We then load bytes at an offset of 1 from this position, i.e., \texttt{41}--\texttt{11} into another vector register $V_2$. We repeat this process for a total of \textit{SeqLength} vector registers, i.e., until $V_3$ in this case. 

In Step 2, we run a vector comparison operation \texttt{mm512\_cmpgt} between $V_2$ and $V_1$. This operation compares pairwise bytes in both registers to see if the byte from $V_2$ is greater than its counterpart from $V_1$. For instance, byte \texttt{41} from $V_2$ is compared against byte \texttt{AF} from $V_1$ as they are both the first bytes in their respective registers. The operation generates a 64-bit mask $M_1$, containing set bits for all positions where the byte from $V_2$ is greater than that of $V_1$. We repeat this operation between registers $V_3$ and $V_2$ as well to generate mask $M_2$.

In Step 3, we run a single vector comparison operation between registers $V_2$ and $V_1$. This is a \texttt{mm512\_cmplt}, which compares pairwise bytes, checking for bytes in $V_2$ \textit{lesser} than those in $V_1$. The operation generates a bit mask $M_3$. Note that this is an opposite comparison operation to Step 2.

Following this, we check for boundaries and opposing byte pairs (detailed below) before moving the scan position by 64 bytes, i.e., to byte \texttt{11}. Effectively, we are scanning for chunk boundaries and content-defined skips 64 bytes at a time. Finally, note that while trailing bytes \texttt{11} and \texttt{A4} are used in Steps 1 and 2, we have not scanned for boundary sequences beginning at these bytes yet. Thus, they are used again when scanning moves ahead.

\textbf{Boundary Detection.} To detect boundaries, we use the masks obtained in Step 2. Each of these masks contains set bits corresponding to increasing byte pairs. If we combine all the masks using a bitwise \texttt{AND} operation, the resulting mask only contains set bits in positions with increasing bytes from all pairwise vector comparisons. If a bit at index \textit{k} is set within $M_1$, it indicates that the byte at index \textit{k} in $V_2$ is greater than the one at index \textit{k} from $V_1$. Similarly, a set bit index \textit{k} in $M_2$ indicates that the byte at index \textit{k} in $V_3$ is greater than the byte at index \textit{k} from $V_2$. 

For instance, in Figure \ref{fig:seqcdc_vector}, the bit at index \textit{61} in mask $M_1$ represents a comparison between byte \texttt{13} from $V_2$ and byte \texttt{12} from $V_1$ and will have a set bit. Similarly, the bit with index \textit{61} in $M_2$ will be set as it compares byte \texttt{14} from $V_3$ with byte \texttt{13} from $V_2$.  Thus, the resulting mask obtained using $M_1$\texttt{\&}$M_2$ will have a set bit at index \textit{61}.

Boundaries can be detected by examining the resulting combined mask $M_1$\texttt{\&}$M_2$. If the mask contains any set bits (has a non-zero value), a chunk boundary is declared at \textit{SeqLength} bytes ahead of the first set bit's position.

\textbf{Content-defined skipping.} To detect opposing pairs of bytes, we use mask $M_3$ obtained in Step 3. This mask contains set bits at all positions where a byte from $V_2$ is lesser than its counterpart from $V_1$. Thus, the total opposing byte pairs observed in the current scanned region equals the number of set bits in $M_3$. We keep a running total of the number of these opposing byte pairs. When the running total exceeds \textit{SkipTrigger}, a content-defined skip of \textit{SkipSize} bytes is initiated as described in \S\ref{sec:design-skip}. The exact position to jump from is determined using the first set bit that causes the total to exceed \textit{SkipTrigger}.

\textbf{Accelerated x86 Intrinsics.} Intel and AMD CPUs support many other hardware-accelerated intrinsics. Forward Scan (\texttt{builtin\_ffs}) is used to find the first set bit in 32/64-bit integers. Parallel Bit Deposit (\texttt{pdep}) is used to deposit contiguous low bits into a destination integer. Trailing Zero Count (\texttt{tzcnt}) is used to count the number of trailing zeros in an integer. Population Count (\texttt{popcnt}) is used to count the number of set bits in an integer. While these are hardware accelerated, they fall under other CPU instruction sets i.e., they are not vector instructions.

Boundary detection needs to identify the first set bit in a mask, while content-defined skips need to identify the first set bit that causes the running total to exceed \textit{SkipTrigger}, i.e., the $n^{th}$ set bit. We use hardware-accelerated x86 intrinsics \cite{intelIntrinsics} for both of these; \texttt{builtin\_ffs} for boundary detection, and a combination of \texttt{pdep} and \texttt{tzcnt} for content-defined skipping. In addition, we use \texttt{popcnt} within content-defined skipping to count the total number of set bits in the opposing slope mask. Note that the performance of these intrinsics varies across CPU architectures.

\section{Impact of Insertions and deletions}
\label{sec:byte shifting}

\begin{figure}[t]
\centering
    \begin{subfigure}[t]{0.99\linewidth}
        \centering\includegraphics[width=\linewidth]{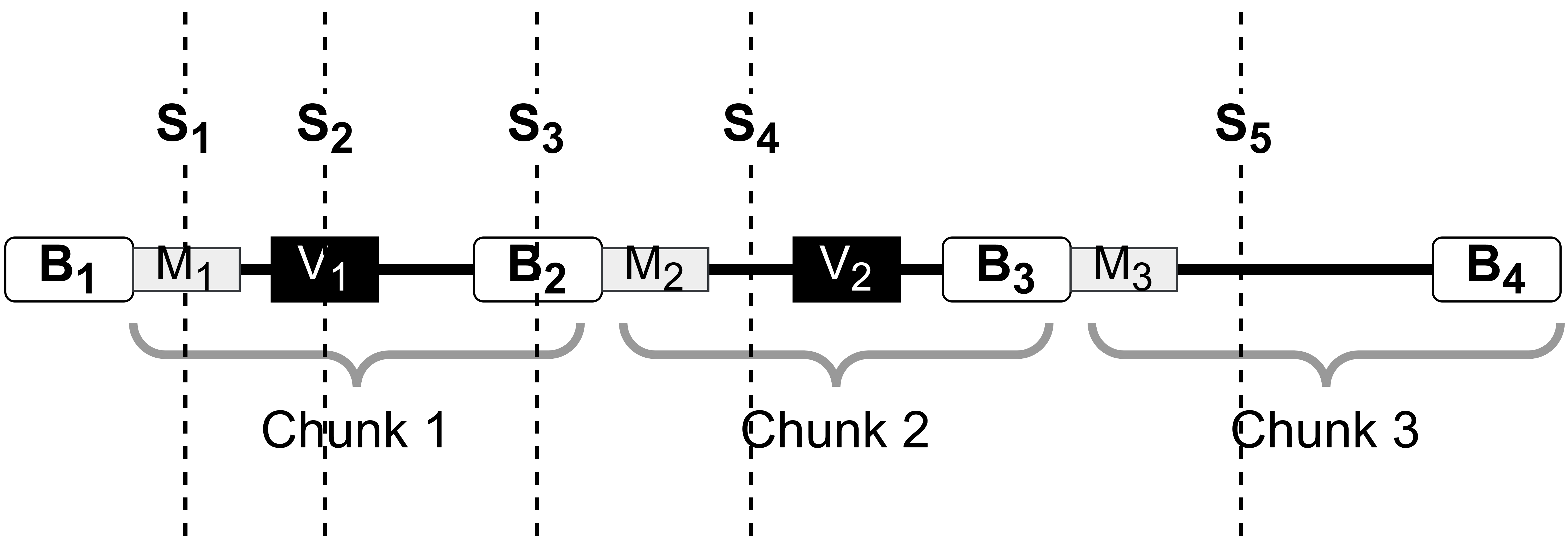}
    \end{subfigure}
    \caption{Handling byte-shifts with \sysname}
    \label{fig:byteshift}
\end{figure}

Inserting or deleting bytes from the middle of a file causes the data bytes to shift, resulting in changes to certain chunk boundaries. Byte shifting can span one or more bytes and take the form of insertions or deletions. In general, sub-minimum and content-defined skipped regions affect \sysname's byte-shifting resistance. Figure \ref{fig:byteshift} shows three chunks with four boundary sequences $B_1$ - $B_4$. Each chunk has a corresponding sub-minimum region ($M_1$ - $M_3$) at the beginning of the chunk. The figure also shows two regions $V_1$ and $V_2$ skipped via \sysname's content-defined skipping i.e., using \textit{SkipTrigger}, and five byte shifts $S_1 - S_5$.

\textbf{Skipped region impact:} When byte-shifts occur, boundaries may be moved in and out of skipped regions (both sub-minimum and content-defined). This occurs with large byte-shifts when boundaries are close to these regions. For instance, consider $S_1$ in Figure \ref{fig:byteshift} which occurs in the sub-minimum region $M_1$. If byte-shift $S_1$ causes a boundary previously hidden within $M_1$ to be pushed outside, a new chunk may be created, thus splitting Chunk 1. This may in turn lead $B_2$ to be pushed into $M_2$, affecting a few subsequent chunks as well. Similarly, a deletion may cause $B_2$ to be hidden within $M_1$, leading to chunk mergers. Additionally, boundaries may be moved in and out of regions previously skipped with \textit{SkipTrigger} and \textit{SkipSize} by shifts such as $S_2$, causing chunk splits and mergers.

While this behavior can theoretically impact a large number of chunks, it only impacts a limited number of chunks within real datasets. This is why many CDC algorithms, such as FastCDC \cite{fastcdc} and TTTD \cite{tttd} use sub-minimum skips in production. \sysname also minimizes the impact caused by content-defined skipping by keeping the \textit{SkipSize} between 256-512 bytes. Thus, despite data skipping, \sysname achieves competitive space savings with other CDC algorithms, as shown in \S\ref{sec:eval_space_savings}. Finally, we note that all CDC algorithms have pathological data patterns i.e. data engineered to ensure that they are ineffective. 

In the rest of this section, we focus on the more common kind of byte-shifting i.e. those that do not result in new boundaries being uncovered or hidden, simply being shifted instead.

\textbf{Within the boundary sequence.} The probability of byte shifting occurring within a boundary sequence is rare in real datasets, as \textit{SeqLength} typically ranges from 3 to 7 bytes. If byte shifting occurs within a sequence, the boundary may no longer exist. Thus, scanning will continue until the next sequence is detected. In Figure \ref{fig:byteshift}, $S_3$ may cause $B_2$ to no longer exist. Thus, the next boundary will be after $B_3$, causing Chunks 1 and 2 to be merged while subsequent chunks are unaffected.

\textbf{Between sequences and outside skipped regions.} Byte-shifts such as $S_4$ are the ones most commonly seen in real datasets. They occur outside of skipped regions and do not drastically change the chunk structure. If $S_4$ does not create a new boundary sequence, the existing boundary sequence $B_3$ shifts. Thus, Chunk 2 changes while subsequent chunks are unaffected. On the other hand, if $S_4$ does create a new boundary sequence, a boundary is inserted after it, changing Chunk 2. Depending on the shift position, the next boundary detected may be $B_3$ or $B_4$, resulting in a changed Chunk 3 as well. Thus, Chunks 2 and 3 are affected while others are unaffected. 

\textbf{Maximum chunk size.} If a shift $S_5$ causes the maximum chunk size to be reached, a boundary is inserted at the maximum chunk size similar to existing algorithms \cite{ae, ram, fastcdc}. This will cause the chunk to split into multiple chunks. For instance, if a boundary is inserted after $S_5$ in Figure \ref{fig:byteshift}, Chunk 3 will be split into two if no subsequent boundary sequences are hidden.

\section{Implementation}
\label{sec:implementation}

{\renewcommand{\arraystretch}{1.25}%
    \begin{table}[t]
    \small
    \centering
        \begin{tabular}{|c|c|c|c|c|}
            \hline
            \rowcolor[HTML]{000000} 
                          & {\color[HTML]{FFFFFF} \textit{\textbf{SeqLength}}} & {\color[HTML]{FFFFFF} \textit{\textbf{SkipTrigger}}} & {\color[HTML]{FFFFFF} \textit{\textbf{SkipSize}}} \\ \hline
            \textbf{4\,KB}                                       & 5                                                  & 55                                                   & 256                                               \\ \hline
            \textbf{8\,KB}                                        & 5                                                  & 50                                                   & 256                                               \\ \hline
            \textbf{16\,KB}                                       & 5                                                  & 50                                                   & 512                                               \\ \hline
        \end{tabular}
    \caption{\sysname parameter values}
    \label{tbl:param_vals}
    \end{table}
}

We implemented a native (unaccelerated) version of \sysname using approximately 250 lines of C++ code. We optimized the computation of \textit{SeqLength} and \textit{SkipTrigger} using the \texttt{std::signbit} function to reduce the number of branch conditions. We have released this code publicly with DedupBench~\cite{dedupbench}.

We implemented the vector acceleration described in \S\ref{sec:design-vector} using an additional 350 lines of C++ code. We have implemented SSE-128, AVX-256 and AVX-512 compatible accelerations for \sysname. \sysname is compatible with all file and data formats.



\textbf{Obtaining parameter values}. \sysname has three configurable parameters apart from its mode: \textit{SeqLength}, \textit{SkipTrigger} and \textit{SkipSize}. To obtain the parameter value combination needed to generate a given average chunk size, we first used Monte-Carlo simulations~\cite{mooney1997monte} on randomized data streams. For example, to identify the parameter values for an average of 4\,KB, we conducted simulations on randomized data to identify candidate combinations resulting in an average chunk size of 4\,KB. Following this, we experimented on one of our datasets (\texttt{DEB} from \S\ref{sec:evaluation}) to pick the best performing combination, which we use across all the datasets in our evaluation. The final values used within our evaluation (\S\ref{sec:evaluation}) for average chunk sizes of 4-16\,KB are shown in Table \ref{tbl:param_vals}.

The results outlined in \S\ref{sec:evaluation} show that \sysname outperforms other CDC algorithms on all datasets with these chosen parameters. Thus, it is not necessary to run an extensive parameter search per dataset, instead obtaining suitable parameters using the method described above. However, to tune \sysname to a specific dataset, better parameter combinations may be obtained using such a search.

\textbf{x86 intrinsics performance.} As noted in \S\ref{sec:design-vector}, the performance of \texttt{popcnt}, \texttt{pdep} and \texttt{tzcnt} may vary across CPU architectures. In order to estimate their performance on a given CPU, we implemented an x86 intrinsics micro benchmark using approximately 250 lines of C++ code. We will release this micro benchmark with the final version of our paper.

 




\section{Evaluation}
\label{sec:evaluation}

\begin{table}[t]
\centering
\small
\begin{tabular}{@{}|c|c|c|c|@{}}
\rowcolor[HTML]{000000} 
{\color[HTML]{FFFFFF} \textbf{Dataset}} & {\color[HTML]{FFFFFF} \textbf{Size}} & {\color[HTML]{FFFFFF} \textbf{Information}}                                                                       & {\color[HTML]{FFFFFF} \textbf{XC}} \\
\textbf{DEB}                                  & 40\,GB                                 & \begin{tabular}[c]{@{}c@{}}65 Debian VM Images from\\ VMware Marketplace \cite{vmware}\end{tabular}                       & 18.98\%                                \\
\textbf{DEV}                                  & 230\,GB                                & \begin{tabular}[c]{@{}c@{}}100 backups of a Rust \cite{github_rust}\\nightly build server\end{tabular}                                                                        & 83.17\%                                \\
\textbf{LNX}                                 & 65\,GB                                 & \begin{tabular}[c]{@{}c@{}}160 Linux kernel distributions\\ in TAR format \cite{kernel_linux}\end{tabular}                   & 19.87\%                                \\
\textbf{RDS}                                  & 122\,GB                                & \begin{tabular}[c]{@{}c@{}}100 Redis \cite{redis} snapshots with \\\texttt{redis-}\texttt{benchmark} runs\end{tabular}        & 33.54\%                                \\
\textbf{TPCC }                                   & 106\,GB        & \begin{tabular}[c]{@{}c@{}}25 snapshots of a MySQL \cite{mysql} \\   VM running TPC-C \cite{TPCCOverview}.\end{tabular} & 37.39\% \\\hline                              
\end{tabular}
\caption{Dataset information. Note that \textit{XC} is the space savings achieved using 8~KB fixed-size chunks.}
\label{tbl:dataset_info}
\vspace*{-\baselineskip}
\end{table}


In this section, we evaluate \sysname's space savings, chunk size distribution, and chunking throughput and compare it to the state-of-the-art CDC algorithms.

\textbf{Testbed.} We use machines with Intel Skylake and Icelake CPUs from CloudLab Wisconsin~\cite{cloudlab_paper} for our evaluation. The Skylake machine (\textit{c220g5}) consists of two 10-core Intel Xeon Silver 4114 CPUs with hyperthreading, 192 GB of RAM, and a 10 GBps Intel NIC, The Icelake machine (\textit{sm220u}) consists of two 16-core Intel Xeon Silver 4314 CPUs with hyperthreading, 256 GB of RAM and a 100 GBps Mellanox NIC. Unless otherwise mentioned, we show results from the Icelake machine. All our results are the average of 5 runs with a standard deviation of less than 5\%.

\textbf{Alternatives.} We evaluate the following hash-based CDC algorithms:
\begin{itemize}
\setlength\itemsep{0.5em}
    \item \textit{CRC}: A hash-based chunking algorithm using CRC-32 \cite{sscdc}.
    \item \textit{FCDC}: FastCDC~\cite{fastcdc} with a normalization level of 2. 
    \item \textit{GEAR}: Gear-based chunking \cite{gear_hash}.
    \item \textit{RC}: A hash-based CDC using Rabin's fingerprinting algorithm~\cite{lbfs}.
    \item \textit{SS-CRC}: AVX-512 version of CRC, accelerated with SS-CDC \cite{sscdc}.
    \item \textit{SS-Gear}: AVX-512 version of Gear, accelerated with SS-CDC \cite{sscdc}.
    \item \textit{TTTD}: Two-Threshold Two-Divisor Algorithm, based on Rabin's fingerprinting with a backup divisor~\cite{tttd_hp}.
\end{itemize}

We also evaluate the following hashless CDC algorithms:
\begin{itemize}
\setlength \itemsep{0.5em}
    \item \textit{AE}: The Asymmetric Extremum~\cite{ae} algorithm.
    \item \textit{RAM}: Rapid Asymmetric Maximum~\cite{ram}.
    \item \textit{VRAM}: SSE-128, AVX-256, and AVX-512 versions of RAM accelerated with VectorCDC \cite{vectorcdc}.
    \item \textit{SEQ}: Native version of \sysname. We only report the results for \sysname in \texttt{Inc\-reas\-ing} mode. The results for \texttt{Decreasing} mode are similar.
    \item \textit{VSEQ}: SSE-128, AVX-256, and AVX-512 versions of \sysname in \texttt{Increasing} mode.
\end{itemize}

We use minimum and maximum chunk sizes of $\frac{1}{2}\times$ and $2\times$ the expected average chunk size, in line with previous studies~\cite{tttd, fastcdc}. The only exception is that for a small average chunk size of 4 KB, we use a minimum size of 1 KB.

\begin{figure}[t]
    \centering
    \hspace*{\fill}
     \begin{subfigure}[b]{0.99\linewidth}
        \centering
        \includegraphics[width=\linewidth]{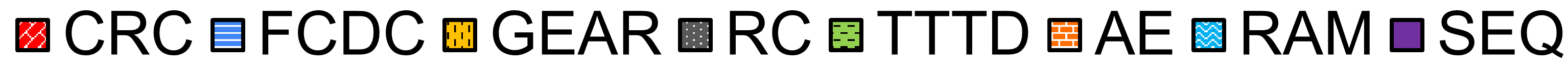}
    \end{subfigure}
    \hspace*{\fill}
    
    \begin{subfigure}[b]{0.49\linewidth}
        \includesvg[inkscapelatex=false,width=\linewidth]{figures/seq-space-deb.svg}
        \caption{\texttt{DEB}}
         \label{fig:space_savings_deb}
    \end{subfigure}
    \begin{subfigure}[b]{0.49\linewidth}
        \includesvg[inkscapelatex=false,width=\linewidth]{figures/seq-space-dev.svg}
        \caption{\texttt{DEV}}
         \label{fig:space_savings_dev}
    \end{subfigure}
    \begin{subfigure}[b]{0.49\linewidth}
        \includesvg[inkscapelatex=false,width=\linewidth]{figures/seq-space-lnx.svg}
        \caption{\texttt{LNX}}
        \label{fig:space_savings_lnx}
    \end{subfigure}   
    \begin{subfigure}[b]{0.49\linewidth}
        \includesvg[inkscapelatex=false,width=\linewidth]{figures/seq-space-rds.svg}
        \caption{\texttt{RDS}}
        \label{fig:space_savings_rds}
    \end{subfigure}
     \begin{subfigure}[b]{0.49\linewidth}
        \includesvg[inkscapelatex=false,width=\linewidth]{figures/seq-space-tpcc.svg}
        \caption{\texttt{TPCC}}
        \label{fig:space_savings_tpcc}
    \end{subfigure}
    \caption{Space Savings with 8KB chunks. Note that \textit{SEQ} is unaccelerated \sysname.}
    \label{fig:space_savings}
\end{figure}

\textbf{Datasets.} Table \ref{tbl:dataset_info} shows the datasets used in our evaluation as well as the space savings achieved by using fixed-size chunking (XC) with an average size of 8~KB. The datasets represent diverse workloads such as database backups, VMs, and Linux kernel code. We have made the \texttt{DEB} dataset publicly available\footnote{\url{https://www.kaggle.com/datasets/sreeharshau/vm-deb-fast25}}~\cite{deb_kaggle}.



\subsection{Space Savings}
\label{sec:eval_space_savings}

\begin{figure}[t]
    \centering
    \begin{subfigure}[b]{0.7\linewidth}
    \centering
        \includegraphics[width=\linewidth]{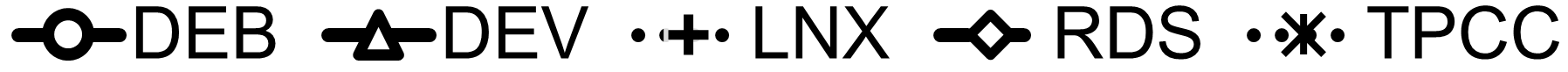}
    \end{subfigure}
    \hspace*{\fill}\\
    \begin{subfigure}[b]{\linewidth}
    \centering
        \includesvg[inkscapelatex=false,width=0.7\linewidth]{figures/seq-space-deg.svg}
    \end{subfigure}   
    \caption{\sysname's Space Savings vs Chunk Size across datasets}
    \label{fig:space_savings_deg}
\end{figure}

Figure \ref{fig:space_savings} shows the space savings achieved by all the alternatives across datasets with 8~KB chunks. By comparing the space savings achieved by fixed-size chunking (\textit{XC}) on all datasets (Table \ref{tbl:dataset_info}) to those achieved by the CDC algorithms (Figure \ref{fig:space_savings}), we note that CDC algorithms achieve superior space savings to \textit{XC}. For instance, \textit{XC} achieves a space savings of only 37.39\% on \texttt{TPCC}, while CDC algorithms achieve 86-87\%. The detailed results for all chunk sizes, algorithms, and dataset combinations are in the Appendix in Table \ref{tbl:space_savings}.


\textbf{CDC comparison.} From Figure \ref{fig:space_savings}, we see that \sysname (\textit{SEQ}) achieves similar space savings to all the other CDC algorithms on these datasets with 8\,KB chunks. The best algorithm for space savings varies depending on the dataset. For instance, \textit{TTTD}, \textit{RAM}, and \textit{SEQ} achieve the best space savings on \texttt{DEV}, \texttt{RDS}, and \texttt{TPCC} respectively. The detailed results for all chunk sizes are in the Appendix in Table \ref{tbl:space_savings}. \textit{On all datasets, \sysname either is the best or achieves space savings within $6\%$ of the best performer for all chunk sizes}.

\textbf{Dataset characteristics.} Figure \ref{fig:space_savings_deg} shows the space savings variation across chunk sizes for all our datasets using \sysname; the results are similar for other CDC algorithms. The space savings achieved decrease with increasing chunk size across datasets. 

The space savings degradation between 4\,KB and 16\,KB average chunk sizes on the \texttt{DEV}, \texttt{RDS}, and \texttt{TPCC} datasets is $0.5$--$6\%$. However, as the total number of chunks at 16\,KB is far lower than that at 4\,KB, the size of the fingerprint database and fingerprinting overheads are significantly lower. 
Similarly, the best chunk size configuration for the \texttt{DEB} dataset is 8\,KB. \textit{This demonstrates why deduplication systems favor larger chunk sizes on some datasets}.

On the other hand, the \texttt{LNX} dataset presents a case favoring smaller chunk sizes. The space savings degradation for \texttt{LNX} in Figure \ref{fig:space_savings_deg} moving from average chunk sizes of 4\,KB to 16\,KB is 29.87\%. This far outweighs any gains within fingerprint indexing.

\textbf{Vector acceleration.} We observed that vector acceleration minimally impacts the space savings achieved by CDC algorithms, in line with previously observed results \cite{sscdc, vectorcdc}. For instance, \textit{SS-CRC} and \textit{CRC} achieve the same space savings values. We have excluded these results from Figure \ref{fig:space_savings} for clarity. Similarly, vector acceleration does not impact \sysname's space savings. 


\begin{figure}[t]
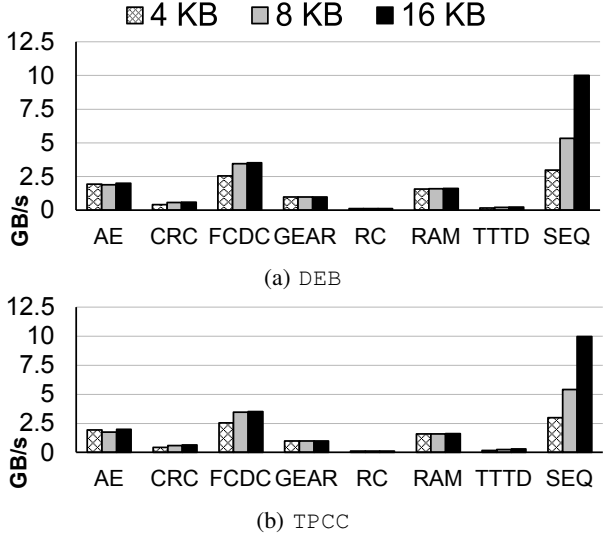

\centering
\begin{subfigure}[b]{0.55\linewidth}
    \centering
    \includegraphics[width=\linewidth]{figures/legend_chunking_speed.png}
\end{subfigure}
\begin{subfigure}[b]{0.9\linewidth}
    \centering
    \includesvg[inkscapelatex=false,width=\linewidth]{figures/debian_throughput.svg}
    \caption{\texttt{DEB}}
\end{subfigure}
\begin{subfigure}[t]{0.9\linewidth}
    \centering
    \includesvg[inkscapelatex=false,width=\linewidth]{figures/tpcc_throughput.svg}
    \caption{\texttt{TPCC}}
    \label{fig:tpcc_throughput}
\end{subfigure}
\caption{Chunking throughput of native CDC algorithms. Note that \textit{SEQ} is unaccelerated \sysname.}
\label{fig:chunking_throughput}
\end{figure}

\subsection{Chunking Throughput: Unaccelerated CDC}
\label{sec:eval_chunking_speed}

Figure \ref{fig:chunking_throughput} shows the chunking throughput for all unaccelerated CDC algorithms on \texttt{DEB} and \texttt{TPCC}. The results for other datasets are similar. When examining chunking throughputs across average chunk sizes, we note that the throughput minimally scales with size for all algorithms other than \textit{SEQ}, similar to our analysis in \S\ref{sec:motivation_chunking_speed}.

Among the hash-based algorithms, \textit{RC}~\cite{lbfs} and \textit{TTTD}~\cite{tttd_hp} are the slowest, only achieving $0.12-0.22$ GB/s. \textit{TTTD} is slightly faster than Rabin's chunking due to sub-minimum skipping (\S\ref{sec:bg_chunking}). The poor chunking throughput of both algorithms is due to the high computational cost of Rabin's hashing, as pointed out in previous literature~\cite{fastcdc}. 
\textit{GEAR} chunking \cite{gear_hash} uses the faster gear hash algorithm allowing it to reach $0.95-0.98$ GB/s. FastCDC (\textit{FCDC}) fares significantly better and achieves $2.5-3.5$ GB/s, largely due to sub-minimum skipping and using the Gear hashing algorithm. \textit{FCDC}'s throughput increases between average chunk sizes of 4\,KB to 8\,KB due to the increased ratio of the sub-minimum region size to chunk size (25\% to 50\%).


Hashless algorithms such as \textit{AE}~\cite{ae} and \textit{RAM}~\cite{ram} achieve higher chunking throughput than all hash-based algorithms except \textit{FCDC}. They both achieve throughputs of $1.5-1.9$\,GB/s across datasets. 


Unaccelerated \sysname (\textit{SEQ}) consistently achieves higher chunking throughput than all other CDC algorithms at average chunk sizes of 8\,KB and 16\,KB. At a chunk size of 8\,KB, it achieves $5.3-5.4$\, GB/s, $1.5\times$ and $2.8\times$ better than \textit{FCDC} and \textit{AE} respectively. At a chunk size of 16\,KB, it achieves a chunking throughput of \textasciitilde$9.9-10$\,GB/s, $2.8\times$ and $5.2\times$ better than \textit{FCDC} and \textit{AE} respectively. \textit{SEQ}'s increase in throughput from 8\,KB to 16\,KB is primarily due to the increasing \textit{SkipSize} from 256 bytes to 512 bytes. 



At a chunk size of 4\,KB, \sysname's \textit{SkipTrigger} is increased by 10\% to constrain the amount of data skipped. 
\sysname still achieves $2.9$\,GB/s, $1.16\times$ and $1.5\times$ faster than \textit{FCDC} and \textit{AE} respectively. Thus, \textit{native \sysname is faster than other native CDC algorithms by $1.5\times$--$2.8\times$ at larger chunk sizes and $16\%$ at smaller chunk sizes}.


\subsection{Chunking Throughput: Vector-accelerated CDC}

\begin{figure}[t]
    \centering
    \begin{subfigure}[t]{.9\linewidth}
        \centering
        \includegraphics[width=\linewidth]{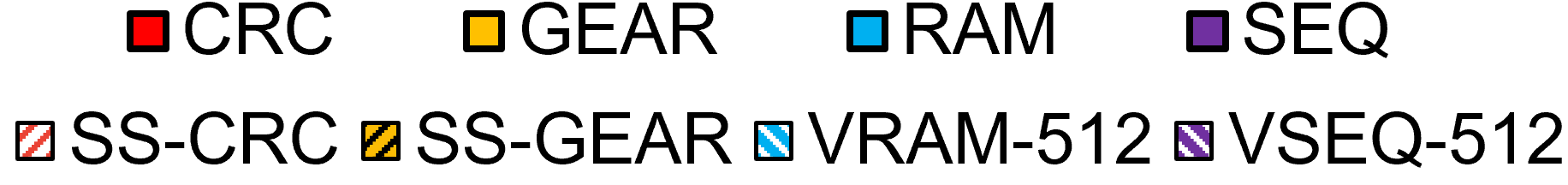}
    \end{subfigure}
    \begin{subfigure}[t]{.49\linewidth}
        \centering
        \includesvg[inkscapelatex=false,width=\linewidth]{figures/debian_throughput_avx_sscdc.svg}
        \caption{CRC and Gear with SS-CDC}
        \label{fig:debian-throughput-speedup-ss}
    \end{subfigure}
    \begin{subfigure}[t]{.49\linewidth}
        \centering
        \includesvg[inkscapelatex=false,width=\linewidth]{figures/debian_throughput_avx.svg}
        \caption{\sysname and VRAM}
        \label{fig:debian-throughput-speedup-seq}
    \end{subfigure}
    \caption{Chunking speedups on \texttt{DEB} with 16\,KB chunks}
    \label{fig:debian_throughput_avx}
\end{figure}

Figure \ref{fig:debian_throughput_avx} shows the throughput of native algorithms with their respective AVX-512 accelerated versions; \textit{CRC} and \textit{Gear} accelerated with SS-CDC \cite{sscdc}, \textit{RAM} accelerated with VectorCDC \cite{vectorcdc} and \sysname accelerated with the method described in \S\ref{sec:design-vector}. Figure \ref{fig:debian-throughput-speedup-seq} shows that \textit{VSEQ-512 achieves the highest throughput among all vector-accelerated algorithms}, at $30.5$\,GB/s. It is also $10\times$ faster than \textit{FCDC}, the fastest non-vector alternative.

\textbf{Vector speedups.} \textit{SS-CRC} and \textit{SS-Gear} (Figure \ref{fig:debian-throughput-speedup-ss}) achieve low speedups over their native counterparts due to using scatter / gather instructions and dependencies between adjacent bytes (\S\ref{sec:bg-vector}). Hashless vector-based algorithms such as \textit{VRAM} and \textit{VSEQ} do not suffer from these limitations. It is important to note that while \textit{VSEQ} achieves the highest throughput overall, its speedup over its native counterpart is lower than that of \textit{VRAM} (Figure \ref{fig:debian-throughput-speedup-seq}). \textit{VSEQ-512} achieves a $3.05\times$ speedup over \textit{SEQ} while \textit{VRAM-512} achieves a $16\times$ speedup over \textit{RAM}. Thus, \textit{\sysname is not as vector friendly as RAM.}

For the rest of our evaluation, we will use only \textit{VSEQ} and \textit{VRAM}.

\begin{figure}[t]
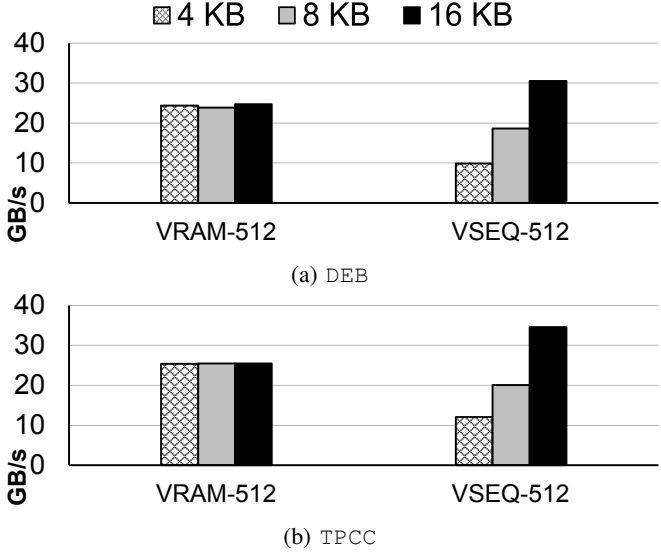

\centering
\begin{subfigure}[b]{0.55\linewidth}
    \centering
    \includegraphics[width=\linewidth]{figures/legend_chunking_speed.png}
\end{subfigure}\\
\begin{subfigure}[b]{0.99\linewidth}
    \centering
    \includesvg[inkscapelatex=false,width=\linewidth]{figures/debian_throughput_vseq_vram.svg}
    \caption{\texttt{DEB}}
    \label{fig:debian_throughput_vseq_vram}
\end{subfigure}
\begin{subfigure}[t]{0.99\linewidth}
    \centering
    \includesvg[inkscapelatex=false,width=\linewidth]{figures/tpcc_throughput_vseq_vram.svg}
    \caption{\texttt{TPCC}}
    \label{fig:tpcc_throughput_vseq_vram}
\end{subfigure}
\caption{Chunking throughput of vector accelerated CDC algorithms. Note that \textit{VSEQ} is vector accelerated \sysname.}
\label{fig:chunking_throughput_vseq_vram_chunksize}
\end{figure}

\textbf{VSEQ vs VRAM.} Figure \ref{fig:chunking_throughput_vseq_vram_chunksize} compares the two best performing alternatives \textit{VSEQ-512} and \textit{VRAM-512} across different chunk sizes on \texttt{DEB} and \texttt{TPCC}. The figure shows that \textit{VRAM} cannot scale its throughput with increasing chunk sizes, similar to its native counterpart \textit{RAM}. Thus, \textit{vector-acceleration alone cannot fix the inability of CDC algorithms to scale their throughput with chunk size}. \textit{VSEQ} can accomplish this due to its changing \textit{SkipSize}, similar to \textit{SEQ}.

At lower chunk sizes such as 4\,KB, we note that \textit{VRAM} outperforms \textit{VSEQ}. However, at 8\,KB this performance gap quickly closes. At 16\,KB, \textit{VSEQ} outperforms \textit{VRAM} by $1.23-1.35\times$. This gap increases further at larger chunk sizes. Thus, while \textit{VRAM} is faster at smaller chunk sizes, \textit{VSEQ is $1.23-1.35\times$ faster than VRAM at the larger chunk sizes favored by deduplication systems.}

\begin{figure}[b]
    \centering
    \begin{subfigure}[t]{.9\linewidth}
        \centering
        \includegraphics[width=\linewidth]{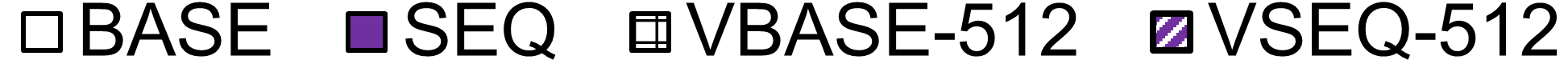}
    \end{subfigure}
    \begin{subfigure}[t]{.9\linewidth}
        \centering
        \includesvg[inkscapelatex=false,width=\linewidth]{figures/speed_breakdown.svg}
    \end{subfigure}
    \caption{\sysname's throughput breakdown at 16\,KB with native and AVX-512 instructions}
    \label{fig:speed_breakdown}
\end{figure}

\textbf{Throughput breakdown.} Figure \ref{fig:speed_breakdown} shows the impact of each optimization (\S \ref{sec:design}) on \sysname's throughput. We use 16\,KB chunks for this experiment. 

\textit{BASE} represents native \sysname running with only lightweight boundary judgement and sub-minimum skipping i.e. without content-defined skipping. \textit{SEQ} enables content-defined skipping. \textit{VBASE-512} represent an AVX-512 accelerated \sysname without content-defined skipping. Finally \textit{VSEQ-512} uses both AVX-512 acceleration and content-defined skips. 

We note that content-defined skipping is beneficial on all datasets without AVX-512 acceleration, albeit to different extents. For instance, content-defined skipping allows \textit{SEQ} to achieve $2.2\times$--$2.6\times$ higher throughput than \textit{BASE} on \texttt{LNX and RDS}. On the other hand, it only achieves  $1.38\times$ on \texttt{DEB}.

\begin{figure}[t]
    \centering
    \begin{subfigure}[b]{0.99\linewidth}
    \centering
        \includegraphics[width=0.9\linewidth]{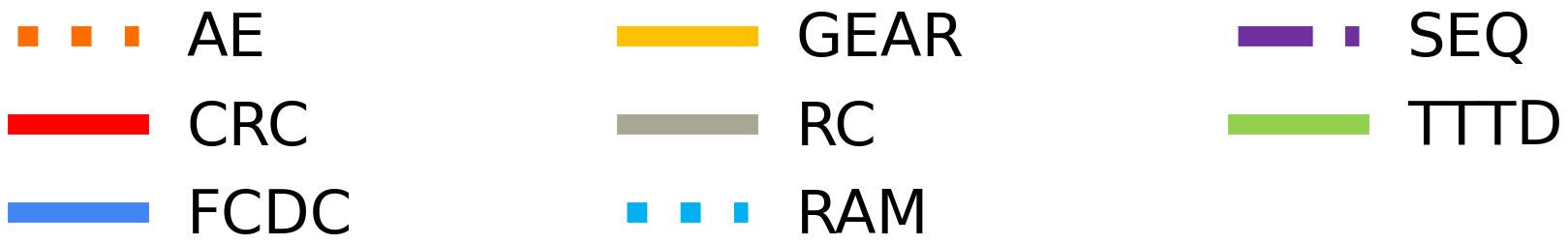}
    \end{subfigure}
    \begin{subfigure}[b]{0.6\linewidth}
    \centering
        \includesvg[inkscapelatex=false,width=\linewidth]{figures/tpcc_8k_distribution.svg}
        \caption{8\,KB}
    \end{subfigure}   
    \begin{subfigure}[b]{0.6\linewidth}
    \centering
        \includesvg[inkscapelatex=false,width=\linewidth]{figures/tpcc_16k_distribution.svg}
        \caption{16\,KB}
    \end{subfigure}
    \caption{Chunk size distribution (CDF) on \texttt{TPCC}}
    \label{fig:chunk_size_distributions}
\end{figure}

When pairing content-defined skips with AVX-512 acceleration, the landscape changes. We notice three kinds of behavior between \textit{VBASE-512} and \textit{VSEQ-512} in Figure \ref{fig:speed_breakdown}. On \texttt{RDS}, the benefits are still significant, with \textit{VSEQ-512} achieving a $2.61\times$ higher throughput than \textit{VBASE-512}. On \texttt{DEV}, \texttt{LNX}, and \texttt{TPCC}, the speedups are slightly lower at $1.12\times$--$1.41\times$. Finally, on \texttt{DEB}, skipping is not beneficial. 

Despite being hardware-accelerated, keeping track of opposing slopes and jump positions does not require vector instructions; it uses Bit Manipulation Instructions (BMI and BMI-2)\cite{haswell_arch}. Such intermingling of vectorized and non-vectorized code regions leads to lower performance gains \cite{avx_overhead}, and may outweigh the benefits of content-defined skips on certain datasets. Thus, while \sysname's \textit{content-defined skipping is almost always beneficial without vector instructions, its benefits depend on dataset characteristics when paired with vector instructions.}


\begin{figure*}[t]
    \centering
    \begin{subfigure}[b]{0.4\linewidth}
    \centering
        \includegraphics[width=0.99\linewidth]{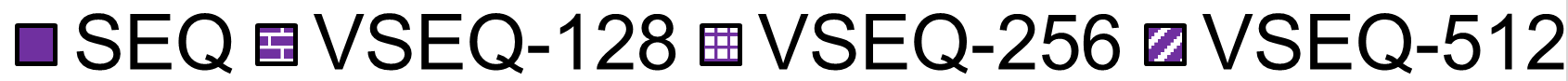}
    \end{subfigure}
     \begin{subfigure}[b]{0.4\linewidth}
    \centering
        \includegraphics[width=0.99\linewidth]{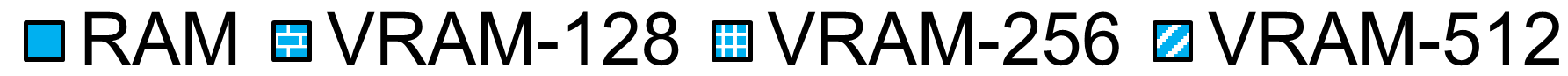}
    \end{subfigure}
    \hfill
    \begin{subfigure}[b]{0.35\linewidth}
    \centering
        \includesvg[inkscapelatex=false,width=\linewidth]{figures/seq-speedup-skylake.svg}
        \caption{\sysname Speedups on Skylake}
        \label{fig:seq-speedup-sl}
    \end{subfigure}
    \begin{subfigure}[b]{0.35\linewidth}
        \centering
        \includesvg[inkscapelatex=false,width=\linewidth]{figures/seq-vram-sse128-skylake.svg}
        \caption{Comparison on Skylake with 16\,KB chunks}
        \label{fig:seq-vram-sl}
    \end{subfigure}   
  
    \begin{subfigure}[b]{0.35\linewidth}
    \centering
        \includesvg[inkscapelatex=false,width=\linewidth]{figures/seq-speedup-icelake.svg}
        \caption{\sysname Speedups on Icelake}
        \label{fig:seq-speedup-il}
    \end{subfigure}
    \begin{subfigure}[b]{0.35\linewidth}
    \centering
        \includesvg[inkscapelatex=false,width=\linewidth]{figures/seq-vram-sse128-icelake.svg}
        \caption{Comparison on Icelake with 16\,KB chunks}
        \label{fig:seq-vram-il}
    \end{subfigure}
    \caption{Chunking Throughput on \texttt{DEB} with SSE-128, AVX-256 and AVX-512 Instructions}
    \label{fig:avx-256}
\end{figure*}

\subsection{Backward compatibility with AVX-256 and SSE-128}

Only a handful number of newer Intel and AMD processors support AVX-512 instructions. However, SSE-128 and AVX-256 instructions have been supported by Intel and AMD since 2003 and 2011 respectively. Thus, most CPUs available today support one of the three classes of vector instructions.

While \sysname's design described in \S\ref{sec:design-vector} uses AVX-512 instructions, \sysname can be accelerated with SSE-128/AVX-256 instructions as well. Figures \ref{fig:seq-speedup-sl} and \ref{fig:seq-speedup-il} show the benefits of accelerating \sysname with SSE-128, AVX-256, and AVX-512 instructions on two different CPU architectures: Intel Skylake and Intel Icelake. We use the \texttt{DEB} dataset.

We note that \textit{VSEQ-128} and \textit{VSEQ-256} achieve significant speedups over \textit{SEQ} on both architectures. For instance with 16\,KB chunks, \textit{VSEQ-128} achieves a $1.57\times$--$2.19\times$ speedup on both architectures. Similarly, \textit{VSEQ-256} at 16\,KB achieves a $2.52\times$--$3.34\times$ speedup. Thus, \textit{\sysname retains most of its throughput benefits when accelerated with SSE-128 and AVX-256 instructions, and is compatible with most modern CPUs}.

Note that the speedups slightly vary across architectures. For instance, \textit{VSEQ-512} achieves a $5.8\times$ speedup over \textit{SEQ} on the Skylake while it achieves $3.05\times$ on the Icelake machine. This is largely tied to CPU microarchitecture and x86 intrinsic performance as described in \S\ref{sec:design-vector}.

\textbf{\sysname vs VRAM.} Figures \ref{fig:seq-vram-sl} and \ref{fig:seq-vram-il}  compare \textit{VSEQ} with the best performing vector accelerated alternative, \textit{VRAM}, using SSE-128, AVX-256 and AVX-512 instructions on Intel Skylake and Icelake architectures. We use \texttt{DEB} with 16\, KB chunks for this experiment. We see that \textit{VSEQ} maintains its performance advantage over \textit{VRAM} with all vector instruction families, further cementing its effectiveness and backward compatibility.

\subsection{Chunk Size Distribution} 

CDC algorithms are expected to achieve a uniform chunk size distribution, loosely centered around the average chunk size. Figure \ref{fig:chunk_size_distributions} shows a CDF of chunk sizes from all algorithms on the \texttt{TPCC} dataset at average sizes of 8\,KB and 16\,KB. Hash-based algorithms' distributions are shown using solid lines, while hashless algorithms use patterned lines. 

Hash-based algorithms exhibit uniform distributions between the minimum and maximum chunk sizes. Rabin's Chunking (\textit{RC}) and \textit{TTTD} exhibit similar distributions since \textit{TTTD} only differs from \textit{RC} by the use of a backup divisor. \textit{CRC} and \textit{GEAR} exhibit similar smooth patterns as well. FastCDC (\textit{FCDC}) is an exception, exhibiting a split pattern centered around the average chunk size. This is because it switches masks and relaxes boundary conditions past the average chunk size i.e., chunk size normalization (\S\ref{sec:bg_chunking}).
\textit{AE} and \textit{RAM} exhibit tighter distributions when compared to hash-based algorithms. 

Despite being hashless, \sysname exhibits a chunk size distribution similar to hash-based algorithms. We observed similar results across all our datasets and chunk sizes.

\textbf{Vector acceleration.} Chunk size distributions remain unaffected by vector acceleration for all CDC algorithms. 


\section{Additional Related Work}
\label{sec:related_work}

\textbf{Accelerating deduplication.} Numerous efforts have been made to accelerate the other phases involved in data deduplication.  Store\-GPU~\cite{storegpu} uses GPUs, Silo \cite{silo} uses locality-based optimizations and other work \cite{balancing_encrypted, optimizing_dedup} focuses on using faster hashing algorithms to improve fingerprint indexing. History-aware rewriting \cite{history_aware_rewrite} and FG-DEFRAG \cite{fgdefrag} improve restore performance by reducing fragmentation in stored chunks. HYDRAstor \cite{hydrastor} and Extreme Binning \cite{extreme_bin} improve scalability by building deduplication on top of distributed storage.

While \sysname is compatible with many of these approaches, they are orthogonal to ours as we focus on the file chunking phase within deduplication.

\textbf{Other chunking optimizations.} RapidCDC \cite{rapidcdc} and QuickCDC~\cite{quickcdc} use locality-based optimizations to speed up chunking for duplicate chunks. MUCH \cite{MUCH} and P-Dedupe \cite{pdedup} parallelize chunking using multiple threads. \sysname is compatible with any of these techniques as they all rely on implementing optimizations on top of existing CDC algorithms. 

MII \cite{mii} uses a sequence-based approach to chunk data but their approach results in inflexible chunk sizes and low throughput. Our previous work \cite{seqcdc_middleware} discussed \sysname briefly. However, it does not present a methodology to accelerate \sysname with vector instructions or compare its performance with other vector-accelerated CDC algorithms.

\textbf{Secure deduplication systems.} Several efforts build end-to-end deduplication systems for encrypted data \cite{secure_dedup_survey}. They mainly target encryption schemes \cite{mle, secure_asokan} for the underlying data or focus on reducing attacks on the system \cite{side_channel_dedup}. \sysname is compatible with all these approaches.


\section{Conclusion}
\label{sec:conclusion}

Deduplication systems in production employ larger chunk sizes due to reduced fingerprinting overheads. However, state-of-the-art CDC algorithms are designed to target smaller average chunk sizes, suffering from poor chunking throughput at larger sizes.

We present \sysname, a CDC algorithm that achieves higher chunking speeds than the state-of-the-art. \sysname leverages hashless lightweight boundary judgement, content-based data skipping, and SSE/AVX acceleration to improve chunking throughput by $10\times$ over unaccelerated and $1.2\times-1.35\times$ over vector accelerated CDC algorithms while achieving similar deduplication space savings. We hope that our work inspires a new generation of vector-friendly data chunking algorithms to accelerate data deduplication.

\section*{Acknowledgments}

We thank Abdelrahman Baba for his contributions to an earlier version of this paper. We thank Lori Paniak for helping us enable our experiments. This research was supported by grants from the National Cybersecurity Consortium (NCC), Natural Sciences and Engineering Research Council of Canada (NSERC) (ALLRP-561423-20 and RGPIN-2025-03332), and Acronis. Sreeharsha is supported by Ontario and Cheriton Graduate Scholarships.

\appendix
Table \ref{tbl:space_savings} shows the space savings achieved by all the alternatives described in \S\ref{sec:evaluation} on all our datasets. The best performing algorithm for each dataset and chunk size combination has been highlighted using bold text. Note that there may sometimes be multiple best-performers.

\sysname (\textit{SEQ}) achieves space savings within 4\% of the best performer on all datasets except \texttt{LNX} and within 6\% of the best performer on \texttt{LNX}.

{\renewcommand{\arraystretch}{1.25}%
\begin{table}[t]
\centering
\small
\begin{tabular}{|c|c|c|c|c|}
\hline
\rowcolor[HTML]{000000} 
{\color[HTML]{FFFFFF} \textbf{Dataset}} &
  {\color[HTML]{FFFFFF} \textbf{CDC}} &
  {\color[HTML]{FFFFFF} \textbf{ 4KB}} &
  {\color[HTML]{FFFFFF} \textbf{ 8KB}} &
  {\color[HTML]{FFFFFF} \textbf{ 16KB}} \\ \hline
  
                 & AE           & 41.99\%          & 33.69\%          & 21.94\%          \\
                 & CRC         & 41.65\%          & 35.23\%          & 26.59\%          \\
                 & FCDC         & 43.83\%          & 36.10\%          & 26.47\%          \\
                 & GEAR         & 39.49\%          & 33.12\%          & 24.12\%          \\
\textbf{DEB}  & RC           & 44.38\%          & 36.16\%          & 27.22\%          \\
                 & RAM          & 42.98\%          & 34.21\%          & 22.37\%          \\
                 & TTTD         & \textbf{45.06\%}          & 37.13\%          & 27.94\%          \\ 
                 & SEQ & 42.77\% & \textbf{37.76\%} & \textbf{29.67\%} \\ \hline
                 
                 & AE           & 98.00\%          & 97.75\%          & 97.21\%          \\
                 & CRC           & 98.15\%          & 98.09\%          & \textbf{97.98\%}          \\
                 & FCDC         & 98.17\%          & 98.06\%          & 97.90\%          \\
                 & GEAR           & 98.14\%          & 98.07\%          & \textbf{97.98\%}          \\
\textbf{DEV}  & RC           & 98.21\%          & 98.12\%          & 97.97\%          \\
                 & RAM          & 98.05\%          & 97.79\%          & 97.31\%          \\
                 & TTTD         & \textbf{98.22\%}         & \textbf{98.12\%}         & \textbf{97.98\%}         \\
                 & SEQ & 98.13\% & 98.03\% & 97.82\% \\ \hline
                 
                 & AE           & 59.41\%          & 45.35\%          & 31.67\%          \\
                 & CRC           & 62.43\%          & 48.70\%          & 36.01\%          \\
                 & FCDC         & 59.16\%          & 43.92\%          & 33.64\%          \\
                 & GEAR           & 57.80\%          & 40.94\%          & 28.93\%          \\
\textbf{LNX} & RC           & 67.02\%          & 50.90\%          & 35.40\%          \\
                 & RAM          & 57.94\%          & 45.62\%          & 29.12\%          \\
                 & TTTD         & \textbf{68.46\%}        & \textbf{51.06\%}         & \textbf{36.92\%}        \\
                 & SEQ & 63.13\% & 49.46\% & 33.26\% \\ \hline
                 
                 & AE           & 94.66\%          & 92.94\%          & 91.04\%          \\
                 & CRC           & 94.66\%          & 93.72\%          & 92.02\%          \\
                 & FCDC         & 93.82\%          & 92.17\%          & 90.15\%          \\
                 & GEAR           & 92.44\%          & 93.01\%          & 89.92\%          \\
\textbf{RDS}  & RC           & 94.31\%          & 92.27\%          & 90.57\%          \\
                 & RAM          & \textbf{95.67\%}        & \textbf{94.09\%}       & \textbf{92.03\%}       \\
                 & TTTD         & 95.2\%           & 92.80\%          & 91.55\%          \\
                 & SEQ & 94.86\% & 92.54\% & 88.78\% \\ \hline
                 
                 & AE           & 86.58\%          & 84.96\%          & 81.58\%          \\
                 & CRC           & 87.23\%          & 86.79\%          & \textbf{86.40\%}          \\
                 & FCDC         & 87.18\%          & 86.74\%          & 86.17\%          \\
                 & GEAR           & 86.96\%          & 86.64\%          & 86.36\%          \\
\textbf{TPCC}    & RC           & 87.24\%          & 86.80\%          & 86.37\%          \\
                 & RAM          & 86.71\%          & 85.21\%          & 81.67\%          \\
                 & TTTD         & \textbf{87.29\%}       & \textbf{86.84\%}        & \textbf{86.40\%}        \\
                 & SEQ & 87.04\% & 86.68\% & 85.83\% \\ \hline
\end{tabular}
\caption{Space savings of CDC techniques. Note that the best algorithm at each configuration is shown using bold text and that \textit{SEQ} is native (unaccelerated) \sysname.}
\label{tbl:space_savings}
\end{table}
}

\bibliographystyle{ieeetr}
\bibliography{references}

\end{document}